\documentclass[12pt]{article}
\usepackage{graphics,cite,epsfig,amsmath,epsf}
\usepackage[english]{babel}
\usepackage{hyperref}
\usepackage{ifpdf}
\usepackage{babel}
\usepackage{multirow}
\usepackage[utf8x]{inputenc} 
\renewcommand{\arraystretch}{1.5}

\newcommand{\lesssim}{ \mathop{}_{\textstyle \sim}^{\textstyle <} }

\begin{document}

\renewcommand{\topfraction}{1.}
\renewcommand{\bottomfraction}{1.}
\renewcommand{\textfraction}{0.}
\renewcommand{\thefootnote}{\fnsymbol{footnote}}
\setcounter{footnote}{0}
\def\thefootnote{\fnsymbol{footnote}}

\begin{titlepage}

\hfill{\tt January 2022}\\

\begin{center}
{\Large \bf Non-Leptonic Decays of Bileptons}

 \bigskip
 \bigskip

 \noindent
{ \bf Gennaro Corcella$^{(a)}$\footnote{gennaro.corcella@lnf.infn.it}, Claudio Corian\`o$^{(b)}$\footnote{claudio.coriano@le.infn.it},\\ Antonio Costantini$^{(c)}$\footnote{antonio.costantini@uclouvain.be}
and Paul H. Frampton$^{(b)}$\footnote{paul.h.frampton@gmail.com}}

\bigskip

{$^{(a)}$ INFN, Laboratori Nazionali di Frascati, \\
Via E. Fermi 54, 00044, Frascati (Rome), Italy.}

{$^{(b)}$ Dipartimento di Matematica e Fisica `Ennio De Giorgi',\\ 
  Universit\`a del Salento, and INFN, Sezione di Lecce,
  Via Arnesano, 73100, Lecce, Italy.}

{$^{(c)}$ Center for Cosmology, Particle Physics and Phenomenology (CP3),\\
  Universit\'e Catholique de Louvain, Chemin du Cyclotron, B--1348,
Louvain la Neuve, Belgium}

\end{center}

\bigskip
\bigskip

\begin{abstract}
We provide a detailed analysis of the 
decays of doubly-charged bilepton gauge bosons $Y^{\pm\pm}$, as predicted in
models based on a $SU(3)_C\times SU(3)_L\times U(1)_X$ symmetry.
In addition to the
decay modes into same-sign lepton pairs, which were already investigated in scenarios wherein
each branching ratio was about one third, there are,
depending on the mass spectrum, possible non-leptonic
decays which reduce the leptonic rates.
These non-leptonic decays
are typically into a light quark (antiquark) and a heavy TeV-scale
antiquark (quark), which carries double lepton
number and decays via virtual bileptons. We choose two benchmark points
to represent the particle spectrum and 
present a phe\-no\-me\-no\-lo\-gi\-cal
analysis of such decays.
We find that, while the LHC
statistics are too low, even in the high-luminosity phase,
they can lead to a visible signal at a future
100 TeV hadron collider (FCC-$hh$). A more exhaustive exploration with a rigorous
inclusion of
tagging efficiencies, detector simulations and higher-order corrections
to the partonic cross section, aiming at assessing the reach of
LHC and FCC-$hh$ for bileptons as a function of the model parameters, 
as well as a recast of four-top searches
in terms of the bilepton model, 
are deferred to future work.
\end{abstract}

\end{titlepage}

\renewcommand{\thepage}{\arabic{page}}
\setcounter{page}{1}
\setcounter{footnote}{0}
\renewcommand{\thefootnote}{\arabic{footnote}}

\newpage

\section{Introduction}

Bileptons appear in the model introduced in \cite{PHF,PP} as
spin-one gauge bosons which have the distinctive properties of
$|Q|=|L|=2$, where $Q$ is electric charge and $L$ the lepton number:
their name became bileptons (from dileptons) in \cite{dilepton}.
In the original paper \cite{PHF} the model was named 331 model,
being based on a $SU(3)_C\times SU(3)_L\times U(1)_X$ symmetry, while
in \cite{CCCF1} the authors have preferred calling it bilepton model,
because it is necessary
to select a well-defined embedding of the electric charge
operator to characterise
adequately the theory, and not only the gauge group.

The advantages of the bilepton model are the explanation for three
quark-lepton families, as well as the
asymmetry of the third quark family, and the necessity that the new
physics must 
lie in the TeV range and cannot be pushed to a higher mass scale.
The principal new particles which are predicted by this
model and will be our concern hereafter are the
gauge bileptons $Y^{\pm\pm}$ and the three extra quarks $D$, $S$ and $T$
with charges $Q = -4/3$, $-4/3$ and $+5/3$, respectively, and lepton number
$L=+2$ ($D$ and $S$) and $-2$ ($T$).

The present paper can be regarded as the third in a series after
\cite{CCCF1,CCCF2} studying bilepton phenomenology.
In fact, the most popular scenarios of physics Beyond the
Standard Model, such as supersymmetry or Universal Extra Dimensions, have
given no signal of new physics at the LHC, and therefore we find it
mandatory investigating alternative scenarios, such as
the bilepton model and the new particles contained in its spectrum.
In detail, 
Ref.~\cite{CCCF1} explored the production at LHC of pairs of gauge bileptons,
decaying into same-sign lepton pairs and
accompanied by two jets. Ref.~\cite{CCCF2} added to the model Lagrangian
an extra Higgs scalar sector, which is a sextet of $SU(3)_L$, leading
to physical doubly-charged Higgs bosons.
The phenomenology of scalar and gauge bileptons was studied in
\cite{CCCF2} in a scenario wherein both decay into same-sign lepton pairs
with branching ratio 1/3 for each lepton species. The final result
was that the bilepton signal could be discriminated from the background,
with the gauge bileptons dominating over the scalars\footnote{See also
  Ref.~\cite{doubly} for an investigation on doubly-charged particles
  with different spin at LHC in the framework of simplified models.}
  
In this paper we shall concentrate on
decays involving the extra quarks, in such a way to complete the exploration
of final states yielded by bileptons and make a final statement on which
signal one should look for at LHC and future colliders.
We shall not account for scalar bileptons: in fact, while
Higgs bosons with charge $\pm 2$ are predicted in several models, doubly-charged
gauge bosons are instead a rather unique feature of the model in \cite{PHF}.
Including doubly-charged Higgs bosons through a $SU(3)_L$ sextet 
as done in \cite{CCCF2} is nevertheless straightforward.

In our analysis we need to estimate the masses of the bilepton
$M_Y$ and of the three quarks $M_D$, $M_S$ and $M_T$.
For $M_Y$ there is a firm
lower limit arising from two different low-energy experiments performed
two decades ago. The first is muonium-antimuonium conversion
$\mu^+e^- \to\mu^-e^+$ which can be mediated by bilepton
exchange \cite{Willmann}; the second is the measurement of the Michel
parameter $\xi$ in muon decay $\mu^- \to e^- \bar\nu_e  \nu_\mu$
which measures \cite{Carlson,Musser,Prieels}
the deviation from pure $(V-A)$ coupling, i.e. $(V -\xi A)$.
The $L$-violating decay 
$\mu^- \to e^- \nu_e + \bar\nu_\mu$ can be mediated
by virtual bilepton exchange and contribute to an effective $(V+A)$
part, hence leading to a departure of $\xi$ below unity. By
coincidence, both experiments agree on a lower mass limit
$M_Y\geq 800$ GeV.
As for high-energy colliders, at the LHC
doubly-charged Higgs bosons were searched
in a number of models by ATLAS \cite{atlashh} and
CMS \cite{cmshh}, under the assumption of a 100\% leptonic rate.
The ATLAS analysis \cite{atlashh}, performed at 13 TeV and with
an integrated luminosity
${\cal L}=36~{\rm fb}^{-1}$, considered the so-called left-right
symmetric model (LRSM) \cite{pati1,pati2}
and its implementation in \cite{spira}, with $H^{\pm\pm}$
coupling to either
left- or right-handed leptons. Ref.~\cite{atlashh}
excluded $H^{\pm\pm}_L$ in the range 770-870 GeV and  $H^{\pm\pm}_R$ in
660-760 GeV. As for CMS, limits between 800 and 820 GeV
were determined at ${\cal L}= 12.9~{\rm fb}^{-1}$ \cite{cmshh}.
More recently, in Ref.~\cite{hhww} the ATLAS Collaboration searched for
doubly-charged Higgs bosons decaying into $W^\pm W^\pm$ pairs at
${\cal L}=139~{\rm fb}^{-1}$, excluding them at 95\% confidence
level up to a mass of 350 GeV.

From a theoretical viewpoint, 
we know from \cite{PHF} that the scale characterizing the symmetry
breaking in the bilepton model is $\lesssim 4$ TeV, which may be
regarded as analogous to the electroweak breaking scale $\langle H\rangle
\sim 248$~GeV. Therefore, since the $W$ mass is about  $\sim \langle
H\rangle/3$, likewise $M_Y\sim 4~{\rm TeV}/3\sim 1.4$~TeV
could be a
reasonable guestimate of the bilepton mass. A more refined and accurate
value of $M_Y$ was nonetheless
obtained in \cite{CF1}, where the authors, by using
renormalization group arguments, gave the estimate $M_Y=(1.29\pm 0.06)$~TeV.

Regarding the new quarks, while
in Refs.~\cite{CCCF1,CCCF2} we assumed that
$D$, $S$ and $T$ were too heavy to contribute to the bilepton width, in this
paper we shall instead
consider benchmarks wherein non-leptonic decays of bileptons are allowed and 
explore the yielded final states at LHC and future 100 TeV proton-proton
collider. In particular, we shall explore a scenario with all such
quarks lighter than $Y^{\pm\pm}$, and another
one wherein only $D$ has mass slightly
below $M_Y$, the others being heavier than bileptons.

A study dealing with the phenomenology of TeV-scale quarks $T$
and bileptons was recently carried out in
\cite{ccgpps}.
However, that analysis is somehow complementary to the present
one: the authors of \cite{ccgpps} explore a scenario with
$T$ heavier than doubly- and singly-charged bileptons $Y^{\pm\pm}$ and $Y^\pm$
and,
rather than bilepton decays into heavy quarks,
as we shall do hereafter, they explore the
decays of $T$ into states with  $Y^{\pm\pm}$ and $Y^\pm$,
first in a simplified-model framework, recasting the CMS analysis
\cite{recast}, and then in the model of Ref.~\cite{PHF}.

Exotic quarks were of course searched at the LHC:
in Ref.~\cite{recast} the CMS collaboration
set the limit $M_T>1.3$~TeV on the mass of heavy quarks $T$ with charge 5/3,
while analyses by ATLAS, such as Ref.~\cite{atlas53}, led to the bound
$M_T>1.2$~TeV. However, these searches
are not directly applicable to our investigation:
our TeV-scale quarks carry lepton number $L=2$ and therefore lead to
final states pretty different from those explored in
\cite{recast,atlas53} for quarks with $L=0$. Nevertheless, a study on 
the primary production of TeV-scale quarks in the bilepton model and the
related phenomenology at LHC and future colliders
is currently under way \cite{CCCF4}.

As in Refs.~\cite{CCCF1, CCCF2}, our exploration will account for two
reference points, which satisfy the theoretical bounds on the model, as
well as the experimental exclusion limits, and investigate the sensitivity
of LHC and possibly of the future FCC-$hh$ ($pp$ collisions
at $\sqrt{s}=100$~TeV) to bilepton discovery.
In principle, an exhaustive exploration of the reach for bileptons
at LHC and ultimately at FCC-$hh$, varying all model parameters, would be useful
and desirable. However, we believe that it would me more appropriate
performing this investigation including  effects like detector simulations,
higher-order corrections to the
partonic cross sections and parton distributions,
which were not included in Refs.~\cite{CCCF1,CCCF2}
and will not be taken into account in the present paper.
We decided to postpone this general analysis to future work
\cite{araz}.

The plan of the present paper is the following.
In Section 2 we shall discuss the benchmarks which we choose and the
corresponding decay rates and branching ratios of 
bileptons and new quarks. In Section 3 we will
present a phenomenological analysis on bilepton signals and backgrounds
at LHC and FCC-$hh$ in our representative points.
Finally, Section 4 will contain some concluding
remarks on the analysis.

\section{Bilepton decays into non-leptonic channels}

As discussed in the introduction, this work will deal with
the phenomenology of bileptons
at LHC and FCC-$hh$, taking particular care about the decays into
the new heavy quarks. 
For the sake of brevity, we shall omit
the theoretical
description of the bilepton model, which can be found in the
pioneering work \cite{PHF}, as well as in the more recent analyses
\cite{CCCF1,CCCF2}. We just recall that, as far as new
particles beyond the Standard Model are concerned, 
the model presents four extra neutral scalar Higgs bosons,
besides the Standard Model one, 
for a total of five $h_1$, $\dots$, $h_5$, three pseudoscalars
$a_1$, $a_2$ and $a_3$, four singly-charged 
$h^\pm_1$, $\dots$ $h^\pm_4$ and three doubly-charged $h^{\pm\pm}_1$,
$h^{\pm\pm}_2$ and $h^{\pm\pm}_3$. The doubly-charged
Higgs bosons predicted by the model have lepton number
$L=\pm 2$ (scalar bileptons) and the phenomenology of the
lightest $H^{\pm\pm}$ was investigated in \cite{CCCF2}
assuming that it decays into same-sign lepton pairs with 100\% branching
ratio, namely 1/3 for each lepton flavour.
Furthermore, one has a $Z'$, which is typically leptophobic, singly- 
($Y^\pm$) and doubly-charged ($Y^{\pm\pm}$) gauge bileptons, on
which the present exploration will be mostly concentrated, as well
as quarks with charge $+ 5/3$ ($T$) and $- 4/3$
($D$ and $S$), whose mass will be assumed at the TeV scale.

The goal of this paper is
presenting a study of non-leptonic decays of bileptons
at LHC and FCC-$hh$ in
a couple of representative points of the parameter space,
consistent with
the theoretical framework and leading to a particle spectrum 
not yet excluded by the experimental searches.
A more general exploration, assessing the dependence of the reach of
such accelerators varying all model parameters is currently under way
\cite{araz}.

In order to get an adequate description of the parameter space,
we will adopt one benchmark point where all 
TeV-scale  quarks are lighter
than the bilepton, and another one with only one quark with mass lower than
$M_{Y^{\pm\pm}}$.
In both cases, the mass of  $Y^{\pm\pm}$
will be fixed about the value estimated in
\cite{CF1}. As in our previous work, the representative points will be obtained
after scanning the parameter space by employing the \texttt{SARAH 4.9.3}
\cite{sarah} code
and its UFO interface \cite{ufo}, implementing the latest exclusion
limits on physics beyond the Standard Model.
Details on the scanning procedure can be found in \cite{CCCF2}.

A first benchmark, named BM I hereafter, will feature quarks
$D$, $S$ and $T$ with mass about 1 TeV; a second one, labelled BM II,
will instead have $S$ and $T$ heavier than $Y^{\pm\pm}$ and $D$ with
mass just below $M_Y$.
Limiting ourselves to Higgs and particles beyond the Standard Model,
the benchmark mass spectrum is presented in Table~\ref{bm}, where only
the masses of $D$, $S$ and $T$ vary between BM I and BM II
and the numbers referring to BM II are quoted in
brackets.\footnote{Table~\ref{bm}
  correctly includes four singly-charged Higgs bosons
  $h^\pm_1$, $\dots$, $h^\pm_4$, unlike Refs.~\cite{CCCF1,CCCF2}, where the
  authors omitted in the tables the fourth charged Higgs $h^\pm_4$.} 
\begin{table}
\begin{center}
  \renewcommand{\arraystretch}{1.4}
\begin{tabular}{|c|c|c|c|}
\hline\hline
\multicolumn{4}{|c|}{Benchmark Point BM I (BM II)}\\
\hline
\hline
$M_{h_1}=125.7$  & $M_{h_2}=3213.5$  & $M_{h_3}=5742.4$  &
$M_{h_4}=17272.8$   \\
\hline $M_{h_5}=17348.3$  & 
$M_{a_1}=5741.0$  & $M_{a_2}=17271.3$  & $M_{a_3}=17348.3$   \\
\hline
$M_{h^\pm_1}=2072.1$  & $M_{h^\pm_2}=5741.7$   & $M_{h^\pm_3}=1727.2$  
&  $M_{h^\pm_4}=17348.5$  \\
\hline
$M_{h^{\pm\pm}_1}=5397.2$  & $M_{h^{\pm\pm}_2}=17191.7$   & $M_{h^{\pm\pm}_3}=17348.6$  &\\
\hline
$M_{Y^{\pm\pm}}=1288.9$  & $M_{Y^\pm}=1291.0$   & $M_{Z'}=4765.9$  &\\
\hline
$M_D=1000.0$ (1200.0)   & $M_S=1000.0$ $(1500.0)$  & $M_T=1000.0$
(1500.0)  & \\
\hline
\hline
\end{tabular}
\caption{Particle masses in GeV in benchmark point BM I (BM II).}
\label{bm}
\end{center}
\end{table}
From Table~\ref{bm}, one learns that 
$D$, $S$ and $T$ have mass 1 TeV in BM I, while in BM II 
$D$ is slightly lighter than $Y^{\pm\pm}$, i.e. $M_D =1.2$~TeV,
with the others heavier, i.e. $M_S=M_T=1.5$~TeV.
Moreover, the Standard Model Higgs ($h_1$)
mass is consistent with the LHC observations,
while the masses of doubly- and singly-charged
bileptons ($M_{Y^\pm}$ and $M_{Y^{\pm\pm}}$) are in agreement with the finding of 
Ref.~\cite{CF1}.
The remaining new particles (Higgs bosons and $Z'$) 
have mass of a few TeV or higher, and therefore they do not
contribute to the bilepton phenomenology which we wish to
explore.

Since in this paper we shall study the non-leptonic decays
of $Y^{\pm\pm}$ and the subsequent decay chains,
it is instructive evaluating widths and branching ratios of bileptons and
TeV-scale quarks.
For this purpose, we use the \texttt{MadGraph} \cite{madgraph}
code, which we shall employ in the following for the event simulation too, 
and in particular its \texttt{MadWidth} module \cite{madwidth}, 
at leading order. We obtain:
\begin{equation}\label{decy1}
  {\rm BR}(Y^{++}\to l^+ l^+)\simeq 20.6\%\ ({\rm BM\ I}),\ 32.5\% ({\rm BM\ II}),
\end{equation}
for each lepton flavour
$l=e,\mu,\tau$. As for non-leptonic decays, the branching ratios read:
\begin{equation}\label{decy2}
  {\rm BR}(Y^{++}\to u\bar D, c\bar S, T\bar b)\simeq 12.7\%\ ({\rm BM\ I}),\ 
  {\rm BR}(Y^{++}\to u\bar D)\simeq 2.5\%\  ({\rm BM\ II}).
  \end{equation}
As expected, in BM I one has substantial branching ratios in all three
non-leptonic modes, which clearly lowers the purely leptonic ones.
In BM II $Y^{\pm\pm}$ mostly decays into same-sign lepton pairs and 
it is only the decay with a $D$
quark in the final state which has a small, though non-negligible,
rate. Therefore, one can already envisage that it is the BM I scenario 
the more promising to study non-leptonic decays.
Overall, the total bilepton widths read:
\begin{equation}\label{wy}
  \Gamma(Y^{\pm\pm})\simeq 17.9~{\rm GeV}\ ({\rm BM\ I});\
  \Gamma(Y^{\pm\pm})\simeq 11.4~{\rm GeV}\ ({\rm BM\ II}).
  \end{equation}
The higher bilepton width in BM I is clearly due to the fact that, unlike
BM II, decays into channels with $S$ and $T$ quarks are permitted.

In BM I, the TeV-scale
quarks exhibit three-body decays into a Standard Model quark
and a same-sign lepton pair or a lepton-neutrino pair, 
through a virtual bilepton. In BM II, $S$ and $T$ are heavier
than singly- and doubly-charged bileptons and can therefore decay
into final states with a real $Y^\pm$ or $Y^{\pm\pm}$.
In BM I the branching ratios of $D$ and $S$, charged $-4/3$, neglecting
light-quark and lepton masses, 
are independent of quark and lepton flavours and read:
\begin{equation}\label{decay1}
  {\rm BR} (D (S)\to u (c)  l^-l^-)\simeq
  {\rm BR} (D (S)\to d (s) l^-\nu_l) \simeq 16.7\%\ 
({\rm BM\ I}).
\end{equation}
In BM II, while the $D$ rates are the same as in BM I, i.e. Eq.~(\ref{decay1}), 
$S$ can decay into real bileptons as follows:
\begin{equation}\label{decay2}
  {\rm BR} (S\to c Y^{--})\simeq 50.5\%,\ 
  {\rm BR} (S\to  sY^-)\simeq 49.5\%\ 
({\rm BM\ II}).
\end{equation}    
The decay rates of $T$, which has charge $+5/3$, are instead given by:
\begin{equation}\label{dect1}
  {\rm BR} (T\to b l^+l^+)\simeq 19.4\%,\ 
  {\rm BR} (T\to t l^+\bar \nu_l)\simeq 13.9\%\  ({\rm BM\ I});
\end{equation}
\begin{equation}\label{dect2}
  {\rm BR} (T\to b Y^{++})\simeq 64.6\%,\  
  {\rm BR} (T\to t Y^+)\simeq 35.4\%\  ({\rm BM\ II}).
  \end{equation}
In BM I the quark $T$ decays into a heavy quark ($b$ or $t$) plus
a lepton pair (two same-sign charged leptons or a charged one and a neutrino)
through virtual $Y^{\pm\pm}$ or $Y^\pm$,
with final states with a $b$ quark exhibiting a higher rate.
In BM II $T$ decays almost exclusively
into a real bilepton and a heavy quark, with the mode $bY^{++}$ being the
dominant one.
Furthermore, in BM I all TeV-scale quark have a quite small
width, of the order of ${\cal O}(10^{-3})~{\rm GeV}$: 
\begin{equation}\label{wq1}
  \Gamma(D)\simeq \Gamma (S) \simeq 3.4\times 10^{-3}~{\rm GeV},\
  \Gamma(T)\simeq 3.0\times 10^{-3}~{\rm GeV}\ ({\rm BM\ I}).
  \end{equation}
In BM II, 
the $D$ width is about ${\cal O}(10^{-2}~{\rm GeV})$, while  
$S$ and $T$, being capable of decaying
into states with real bileptons, have a larger width of the order of 1 GeV:
\begin{equation}\label{wq2}
  \Gamma(D)\simeq 1.3\times 10^{-2}~{\rm GeV},\ 
  \Gamma (S) \simeq 1.5~{\rm GeV},\
  \Gamma(T)\simeq 1.1~{\rm GeV}\ ({\rm BM\ II}).
  \end{equation}
Once the benchmarks have been set and the relevant branching ratios computed,
in the next section we shall perform a phenomenological analysis of non-leptonic
decays of bileptons at LHC and future colliders, aiming at assessing whether
the foreseen cross sections and event numbers are sufficiently high for
any signal to be detectable. Whenever this is the case, we shall explore
possible observations which allow the discrimination from the Standard
Model backgrounds.

Before presenting our result, we point out that, as discussed in
our previous paper \cite{CCCF2}, 
the bilepton model is perturbative up to a
scale about 3.5 TeV \cite{dias,martinez}, while the typical scale of
$Y^{\pm\pm} Y^{\pm\pm}$ production is
$2M_{Y^{\pm\pm} Y^{\pm\pm}}$, i.e. about 2.6 TeV in our case.
Therefore, we shall assume
that a perturbative analysis is legitimate in the energy regime of the present
work.

\section{Phenomenological analysis}

In this section we shall investigate the phenomenology of bilepton
non-leptonic decays in our two benchmark points
at the LHC, i.e. $pp$ collisions at
a centre-of-mass energy
$\sqrt{s}=13$~TeV and integrated luminosity
${\cal L}=300~{\rm fb}^{-1}$, and at a
future hadron colliders (FCC-$hh$) with $\sqrt{s}=100$~TeV
and ${\cal L}=3000~{\rm fb}^{-1}$.
An extension of the 13 TeV results to the high-luminosity
LHC (HL-LHC), namely $\sqrt{s}=14$~TeV and  ${\cal L}=3000~{\rm fb}^{-1}$
is straightforward too.

As discussed in the previous section, hereafter we shall use the
\texttt{MadGraph} code in the leading-order approximation,
deferring the inclusion of NLO corrections to future work.
Through our work, we shall employ the NN23LO1 set of LO parton distribution
functions \cite{nnpdf}, while parton showers and hadronization
will be simulated by the \texttt{HERWIG 6} code \cite{herwig}.

For bileptons at 1.29 TeV, the cross sections for pair production
are given by:
\begin{equation}\label{xlhc}
  \sigma(pp\to Y^{++}Y^{--}) \simeq 0.75~{\rm fb}\ ({\rm LHC}, 13~{\rm TeV});
\end{equation}
\begin{equation}\label{hllhc}
  \sigma(pp\to Y^{++}Y^{--}) \simeq 1.12~{\rm fb}\ ({\rm LHC}, 14~{\rm TeV});
\end{equation}
\begin{equation}\label{xfcc}
  \sigma(pp\to Y^{++}Y^{--}) \simeq 393.89~{\rm fb}\ ({\rm FCC-}hh).
  \end{equation}
As easily predictable, the FCC-$hh$ cross section is about 500 and 350
times larger than the LHC one at 13 and 14 TeV, respectively.

\begin{figure}
\centerline{\resizebox{0.44\textwidth}{!}
{\includegraphics{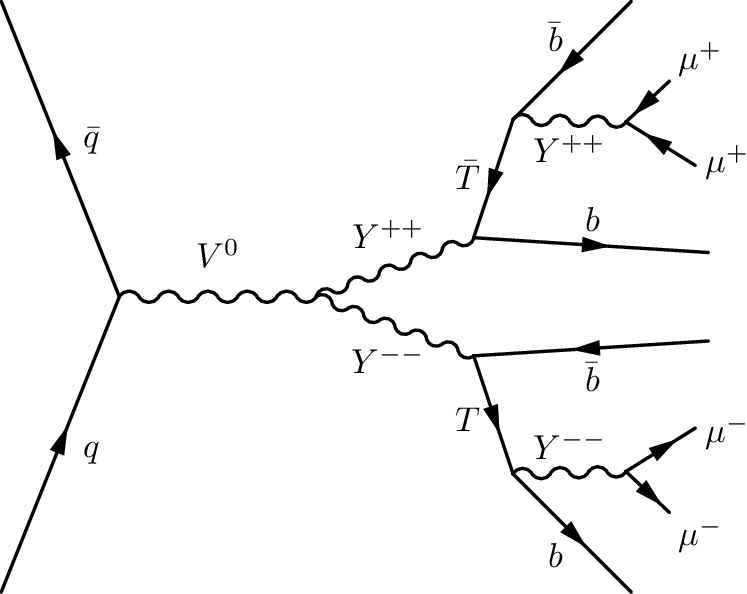}}%
\hfill%
\resizebox{0.44\textwidth}{!}{\includegraphics{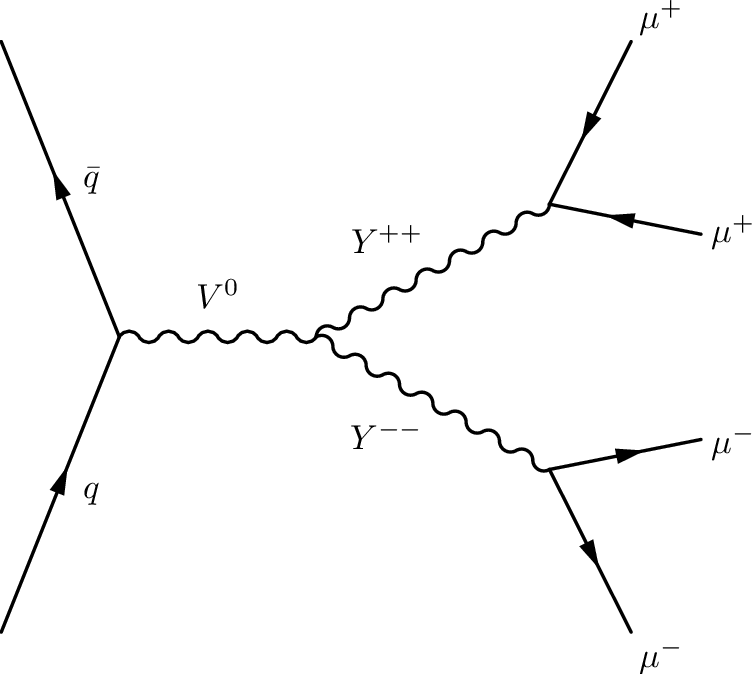}}}
\caption{Left: example of diagram leading to the decay chain 
  (\ref{dec1}), investigated in the present paper.
  Right: Pair-production of bileptons decaying
  into same-sign muons, explored in \cite{CCCF1,CCCF2}.}
\label{diagra}\end{figure}

In the scenario BM I, as a case study, we will investigate
$Y^{\pm\pm}$ primary decays into quarks $T$, which then
decay into a bottom quark and a same sign lepton pair.
Limiting for simplicity our exploration to one lepton species, such as
muons, which have a better charge identification than electrons at LHC
(see, e.g., Refs.~\cite{mu,e}),
this corresponds to the following process\footnote{Of course,
  if we account for both electrons and muons, we will have to
  roughly double cross sections and event rates at LHC and FCC-$hh$.}:
\begin{equation}\label{dec1}
  pp \to Y^{++}Y^{--}\to (T\bar b)(\bar T b)\to (b\bar b \mu^+\mu^+)
(b\bar b\mu^-\mu^-)
  \ ({\rm BM\ I}),\end{equation}
which lead to 
final states with four $b$-flavoured jets and four muons ($4b4\mu$).
In Fig.~\ref{diagra} (left) we display an example of
the decay chain (\ref{dec1}), assuming
a primary $Y^{++}Y^{--}$ production via quark-antiquark annihilation, with the
exchange of a neutral boson $V^0$ (photon, $Z$, $Z'$ or any electrically-neutral
Higgs). For the sake of comparison, in Fig.~\ref{diagra} (right) we also
present a companion diagram, accounted for in \cite{CCCF1,CCCF2}, where
bileptons
decay into same-sign muon pairs. \footnote{We point out that Fig.~\ref{diagra}
shows only a particular contribution to bilepton production and decay
in a $pp$ collision. The \texttt{MadGraph} code calculates
all diagrams allowed by the bilepton model, including,
for example, the production of $Y^{++}Y^{--}$ mediated by the
exchange of a TeV-scale quark in the $t$-channel.}

In reference point BM II, we shall instead analyze
primary decays involving TeV-scale quarks $D$ and yielding
final states with
four $u$-quark initiated light jets accompanied by four muons
($4u4\mu$):
\begin{equation}\label{dec2}
  pp \to Y^{++}Y^{--}\to (\bar D u )(D\bar u)\to (u\bar u \mu^+\mu^+)
(u\bar u\mu^-\mu^-)
  \ ({\rm BM\ II}).\end{equation}
A representative diagram of process (\ref{dec2}) can be obtained
from Fig.~\ref{diagra} (left), after replacing $T$ with $D$ and $b$ with $u$.

An upper limit on the predicted number of
events at LHC and FCC-$hh$
can be estimated by multiplying the inclusive cross sections
in Eqs.~(\ref{xlhc})--(\ref{xfcc}) by the relevant branching ratios,
assuming a perfect tagging efficiency and no cut on final-state
jets and leptons.
Implementing the branching ratios of bileptons and TeV-scale quarks
in Eqs.~(\ref{decy1})--(\ref{dect2}), 
we obtain that the leading-order (LO) 
cross sections 
of the chain (\ref{dec1}) 
at 13 and 14 TeV centre-of-mass energies read:
\begin{equation}
  \sigma(pp\to YY\to 4b4\mu)\simeq 4.55\times 10^{-4}~{\rm fb}\ ({\rm LHC,
    13~TeV,\ BM\ I});
\end{equation}  
\begin{equation}
  \sigma(pp\to YY\to 4b4\mu)\simeq 6.80\times 10^{-4}~{\rm fb}\ ({\rm LHC,
    14~TeV,\ BM\ I}).
\end{equation}  
Such cross sections are
clearly too small to see any event at 300~fb$^{-1}$ and even
at 3000~fb$^{-1}$ (HL-LHC). 
At FCC-$hh$ one instead has:
\begin{equation}
  \sigma(pp\to YY\to 4b4\mu)\simeq 0.24~{\rm fb}\ ({\rm FCC-}hh,\ {\rm BM\ I}),
\end{equation}  
hence a few hundred events can be foreseen at 3000~fb$^{-1}$.
Regarding the BM II scenario, namely the decay chain
(\ref{dec2}), the prediction for the LHC leads to cross sections
of ${\cal O} (10^{-5}~\rm{fb})$ hence
too low for any analysis to be worthwhile
at either 300 or 3000 fb$^{-1}$:
\begin{equation}
  \sigma(pp\to YY\to 4u4\mu)\simeq 1.31\times 10^{-5}~{\rm fb}~({\rm LHC, 13~TeV,\ BM\ II}),
  \end{equation}
\begin{equation}
  \sigma(pp\to YY\to 4u4\mu)\simeq 2.03\times 10^{-5}~{\rm fb}~({\rm LHC, 14~TeV,\ BM\ II}),
  \end{equation}
If one refers instead to FCC-$hh$, the cross section for
the production of four light jets and four muons through the process in
Eq.~(\ref{dec2}) is given by:
\begin{equation}
  \sigma(pp\to YY\to 4u4\mu)\simeq 6.87\times 10^{-3}~{\rm fb}~({\rm FCC}-hh,\
        {\rm BM\ II}).
  \end{equation}
It is therefore a less promising scenario than
BM I even at FCC-$hh$, but it is still
worthwhile checking whether any signal, though small, could be
separated from the background.

Before presenting any phenomenological distribution, 
we need to account for the backgrounds mimicking the bilepton signal,
and set acceptance cuts on both signal and backgrounds.
Because of the expected statistics, we shall limit
ourselves to the FCC-$hh$ case.
As for the chain (\ref{dec1}), leading to four $b$-jets and
four muons, we shall consider the primary
production of four $b$ quarks and two $Z$ bosons decaying into muon pairs
(background $b_1$):
\begin{equation}\label{back1}
  pp\to bb\bar b\bar bZZ\to  bb\bar b\bar b\mu^+\mu^-\mu^+\mu^-.
\end{equation}
We shall also investigate the production
of four top quarks, with the subsequent $W$'s decaying into
muons and requiring, as in \cite{CCCF1}, a small missing energy
due to the muon neutrinos (background $b_2$):
\begin{equation}\label{back2}
  pp\to t t \bar t\bar t\to (bW^+)(bW^+)(\bar bW^-)(\bar bW^-)\to
  bb\bar b \bar b \mu^+\mu^+\mu^-\mu^-\nu_\mu\nu_\mu\bar\nu_\mu\bar\nu_\mu.
\end{equation}
The above background deserves some further comments.
Four-top production occurs mostly through QCD-mediated
processes; however,
it was recently found \cite{zaro} that electroweak contributions,
at both LO and NLO, are non-negligible as they can have an impact
up to 10\% on the total cross section and even larger on
differential distributions. Therefore, we modified the
default four-top \texttt{MadGraph} LO generation, which includes
only QCD interactions, and accounted for the electroweak
contributions as well.
Furthermore, as discussed in \cite{fuks,fuks1}, the analyses on four-top
searches at LHC can be recast to set bounds on several new physics models.
In particular, Ref.~\cite{fuks} recasts the CMS
search \cite{cms4t} at 35.9~fb$^{-1}$ to constrain the sgluon mass, while
Ref.~\cite{fuks1} reinterprets the analysis \cite{cmstttt}
in terms of top-philic simplified-model scalar/vector singlets
or octets (see also \cite{fuks2} for the actual
implementation in the \texttt{MadAnalysis} framework  \cite{madanalysis}).
Such studies tend therefore to suggest that one may even recast the analyses
\cite{cms4t,cmstttt} in terms of the bilepton model at LHC and FCC-$hh$:
this is nevertheless deferred to future work.

Although they are supposed to give a less significant contribution
to the background than processes (\ref{back1}) and (\ref{back2}),
we shall also account for four light jets and two $Z$ bosons decaying
into muon pairs (background $b_3$), i.e.
\begin{equation}\label{back3}
  pp\to jjjjZZ\to jjjj\mu^+\mu^-\mu^+\mu^-.
\end{equation}
as well as processes
with two light jets, two $b$-jets and two $Z\to\mu^+\mu^-$ ($b_4$),
namely 
\begin{equation}\label{back4}
  pp\to jjb\bar bZZ\to jjb\bar b\mu^+\mu^-\mu^+\mu^-,
\end{equation}
where we assume that $j$ is either a light-quark or gluon-initiated jet,
mistagged as a $b$-jet.

Concerning the other scenario, i.e. the decay chain (\ref{dec2}) in
BM II, the main background is the process in Eq.~(\ref{back3}), i.e.
four light jets and
two $Z$ bosons decaying into muons ($b_3$).

In our phenomenological analysis, along the lines of \cite{CCCF1,CCCF2},
we cluster jets according to the $k_T$ algorithm \cite{kt} with
a radius-like parameter
$R=1$ and apply the following acceptance cuts on jets and muons:
\begin{eqnarray}
\label{cuts}
&& p_{T,j}>30~{\rm GeV},\  p_{T,\mu}>20~{\rm GeV},\  |\eta_j|<4.5,
|\eta_\mu|<2.5,\nonumber\\
&& \Delta R_{jj}>0.4,\   \Delta R_{\mu\mu}>0.1,\ 
\Delta R_{j\mu}>0.4,\  {\rm MET}<200~{\rm GeV}.
\end{eqnarray}
As pointed out in \cite{CCCF2}, the cuts in Eq.~(\ref{cuts})
correspond to a conservative choice
of the overlap-removal algorithm implemented to discriminate
lepton and jet tracks at LHC \cite{sidoti,polesello}.
The cut  ${\rm MET}<200$~GeV refers to the missing transverse
energy, i.e.  ${\rm MET}=\sqrt{\left(\sum_{i=\nu}p_{x,i}\right)^2+
  \left(\sum_{i=\nu}p_{y,i}\right)^2}$, due to the neutrinos in the final
state.\footnote{The ${\rm MET}<200$~GeV cut should be applied to the
  neutrinos in $W$ decays in the 4-top background (\ref{back2}) and,
  for the sake of consistency,
  to all neutrinos produced in hadron decays in both
  signal and backgrounds. Nevertheless, we checked that  neutrinos
  from hadron decays are much softer than those in top decays, so that they
  are almost unaffected by the MET cut.}

In principle, for the sake of a reliable analysis, one should
implement the $b$-tagging efficiency, as well as the probability of mistagging
a jet as a $b$-jet.
Also, in principle such efficiencies depend in the jet rapidity and
transverse momentum, as well as 
on the flavour of the parton
which originates the jet.
However, in this explorative letter, in first approximation
we shall implement such effects in a flat manner, i.e. independently
of the jet kinematics, and defer to future work any rigorous
and systematic inclusion of the tagging rates.
Following Ref.~\cite{selvaggi}, we shall assume the following
$b$-tagging efficiency ($\epsilon_b$) and mistag rate ($\epsilon_j$)
\footnote{Strictly speaking, our efficiencies refer to jets with
  pseudorapidity $|\eta_j|<2.5$ and
  transverse momentum $10~{\rm GeV}<p_{T,j}<500~{\rm GeV}$,
  which roughly correspond to the region where most events occur.
  Furthermore, according to \cite{selvaggi}, the mistag rate runs from
  0.01 (jets initiated by $u$, $d$, $s$ quarks or gluons) to
  0.15 (charm-flavoured jets). Our value $\epsilon_j=0.05$ represents
 therefore a sort of weighted average between these two extreme values.}:
\begin{equation}\label{tag}
  \epsilon_b=0.82\ ,\ \epsilon_j=0.05.
  \end{equation}.
Before presenting our results, it is mandatory to calculate the
cross section and the 
number of events foreseen at FCC-$hh$, after the cuts (\ref{cuts})
and the efficiencies (\ref{tag}) are
imposed, for both signal and backgrounds.
Overall, one can envisage that, unlike Refs.~\cite{CCCF1,CCCF2}, where
the signal final states were purely leptonic and all leptons,
coming from the decay of a heavy resonance, had
a sufficiently large transverse momentum to pass the cuts, 
our signal and background processes, involving both jets and
leptons coming from more complex decay chains,
should feel a stronger effect and a more severe suppression
due to the cuts.
From a more technical viewpoint,
we set some basic looser cuts at the level of the matrix-element
generation with \texttt{MadGraph} and the stronger cuts in (\ref{cuts})
after the matching with HERWIG. In fact, without setting
  any cut at parton level, processes like the backgrounds
  (\ref{back1})--(\ref{back2}) would be divergent because of the 
  soft and collinear singularities.  For a review on matching
  matrix elements and parton showers for multi-jet events see
  Ref.~\cite{meps}.


  As for process (\ref{dec1}), we find that, after the acceptance
  cuts are accounted for, the signal ($s$) cross section
  amounts to $\sigma(4b4\mu)_s\simeq 6.24\times 10^{-2}$~fb.
Including the $b$-tagging efficiency as in Eq.~(\ref{tag}), one can then 
envisage $N(4b4\mu)_s\simeq 90$ events
at FCC-$hh$ for a luminosity ${\cal L}=3000$~fb$^{-1}$.
As for the backgrounds (\ref{back1})--(\ref{back4}),
after all cuts, as well as $b$-tagging and
mistag rates are applied, we obtain
$\sigma(4b4\mu)_{b_1}\simeq 1.28\times 10^{-2}$~fb,\\
$\sigma(4b4\mu+{\rm MET})_{b_2}\simeq 3.34\times 10^{-2}$~fb,
$\sigma(4j4\mu)_{b_3}\simeq 4.43$~fb,
$\sigma(2b2j4\mu)_{b_4}\simeq 1.34$~fb.\\
Multiplying such cross sections by 
the luminosity, accounting for the tagging
efficiencies and rounding to the nearest ten, the expected number
of events are then given by 
$N(4b4\mu)_{b_1}\simeq 20$,
$N(4b4\mu+{\rm MET})_{b_2}\simeq 50$, while
backgrounds $b_3$ and $b_4$ yield too few events to be significant 
at FCC-$hh$.

Regarding BM II and the decay chain (\ref{dec2}), the cross section
after the cuts is about $\sigma(4j4\mu)_s\simeq 1.88\times 10^{-3}$~fb, 
FCC-$hh$.
As a result, considering that some extra suppression is to be expected at
FCC-$hh$ due, e.g., to the efficiency of jet/lepton tagging,
which will further lower the event rate of process (\ref{dec2}), 
we prefer to neglect
the BM II scenario and concentrate ourselves on BM I, namely
the decay chain
(\ref{dec1}) and its main backgrounds (\ref{back1}) and (\ref{back2}).

In our phenomenological analysis, we shall investigate the
following observables: the transverse momentum  of the
hardest ($p_{T,1}$) and next-to-hardest ($p_{T,2}$)
muons, the hardest-muon pseudorapidity $\eta_1$, the
invariant mass $M_{\mu\mu}$ of same-sign muons, the polar angle between
same-sign muons $\theta_{\mu\mu}$, the invariant opening angle
$\Delta R_{\mu\mu}=\sqrt{(\Delta\phi)^2+(\Delta\eta)^2}$
between hardest and next-to-hardest muons, $\phi$ being the muon azimuth,
the transverse momentum of the two hardest jets,
$p_{T,j1}$ and $p_{T,j2}$, also named as first and second jet, 
which in our final states are
$b$-flavoured, the invariant opening angle between first jet and hardest
muon ($\Delta R_{j\mu}$) and between the two hardest jets ($\Delta R_{jj}$).

\begin{figure}
\centerline{\resizebox{0.49\textwidth}{!}
{\includegraphics{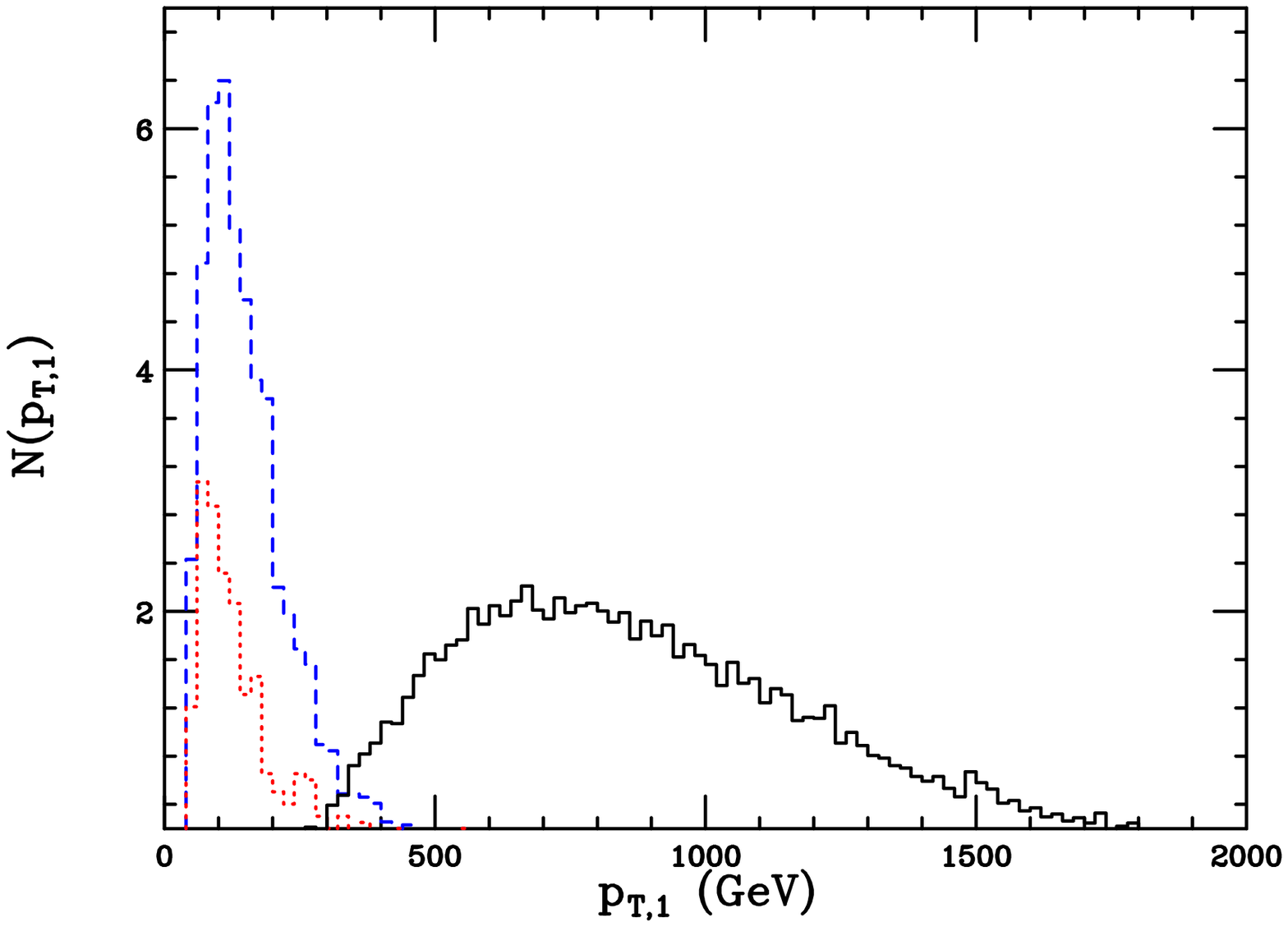}}%
\hfill%
\resizebox{0.49\textwidth}{!}{\includegraphics{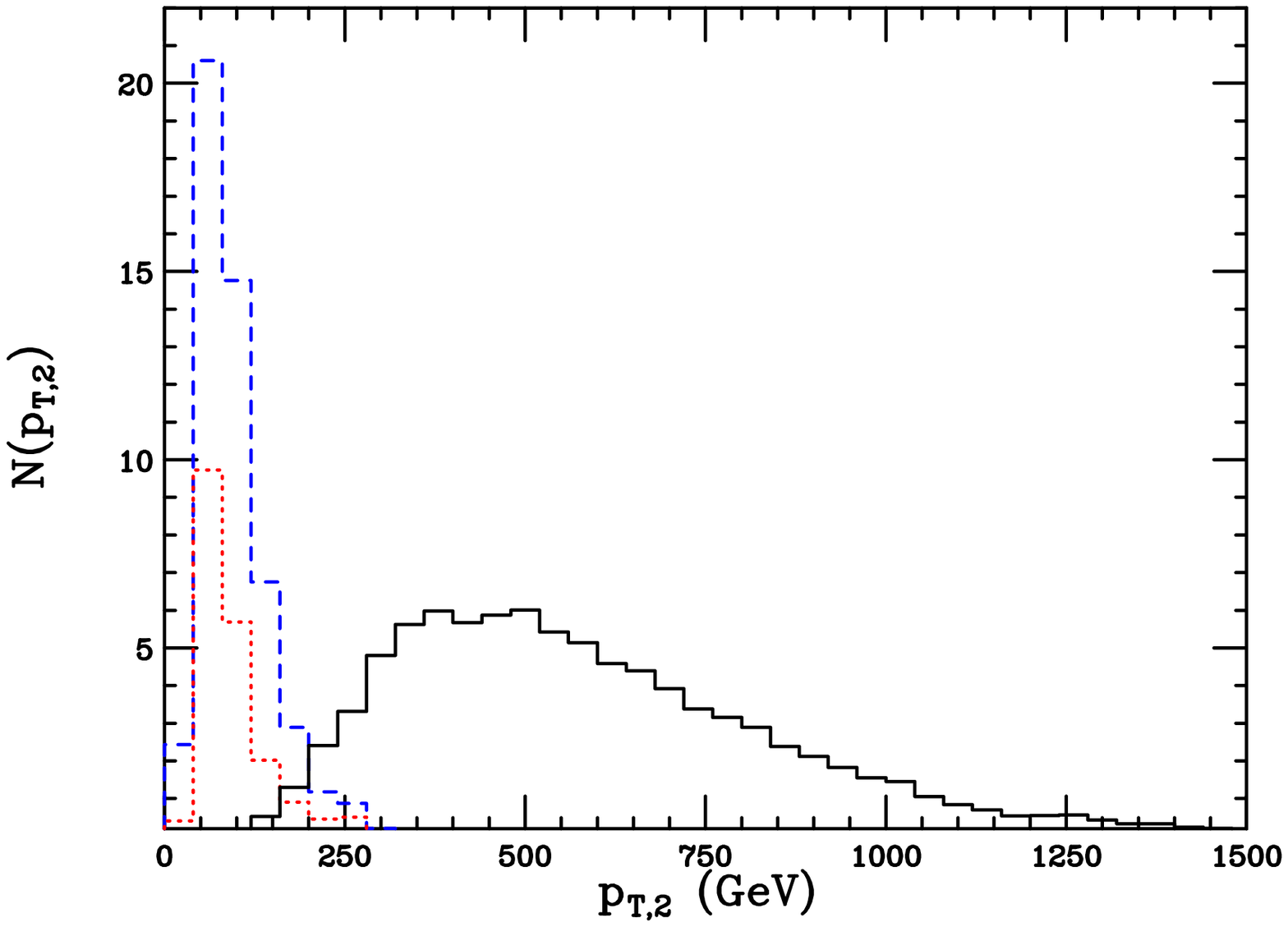}}}
\caption{Transverse momentum of the hardest (left) and next-to-hardest
  muon (right) for the bilepton signal (solid) and the backgrounds
  with four top quarks (dashes) and
  two $b$ quarks and two $Z$ bosons decaying to muons (dots).}
\label{ptlep}
\end{figure}

\begin{figure}
\centerline{\resizebox{0.49\textwidth}{!}
{\includegraphics{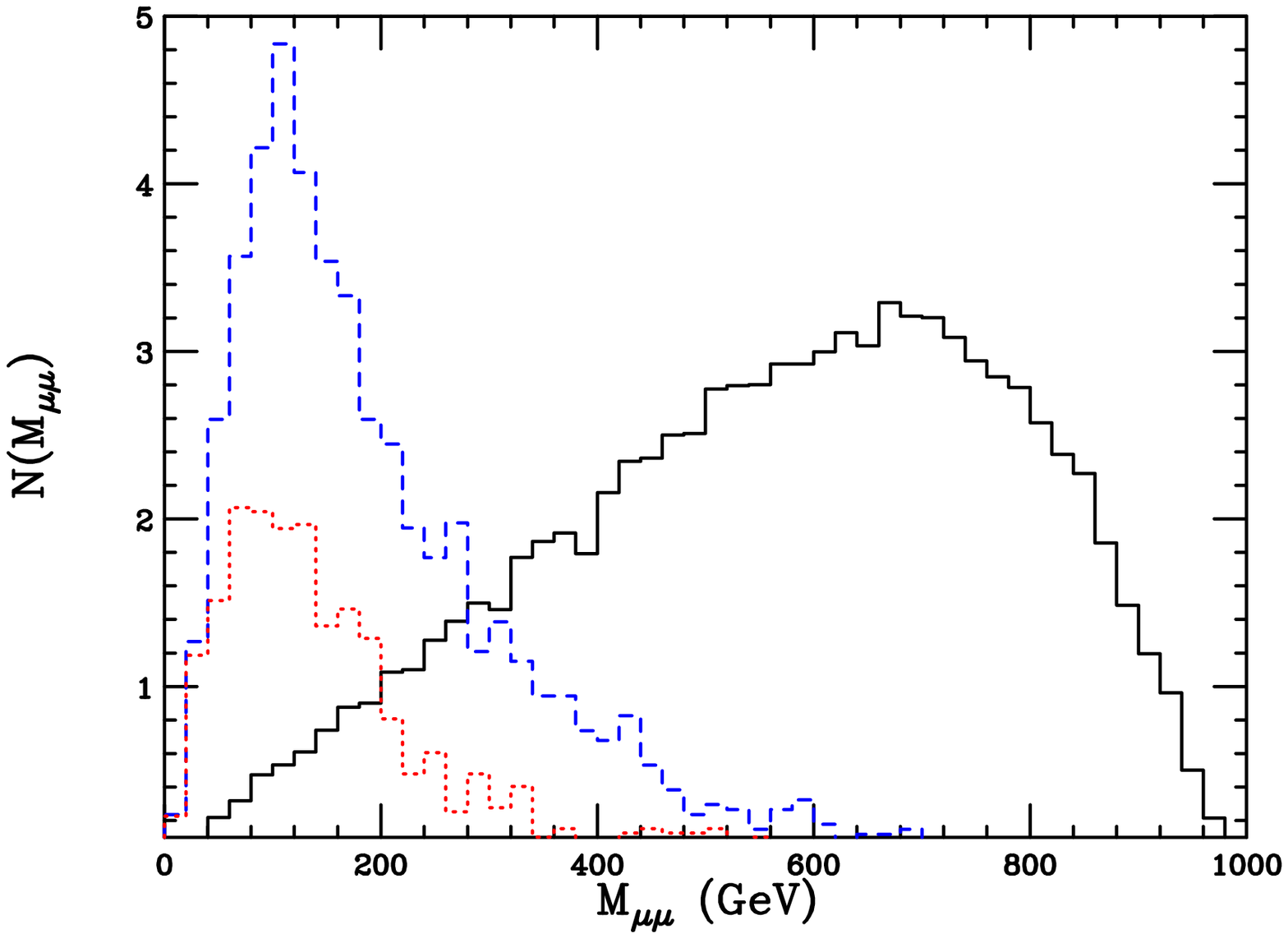}}%
\hfill%
\resizebox{0.49\textwidth}{!}{\includegraphics{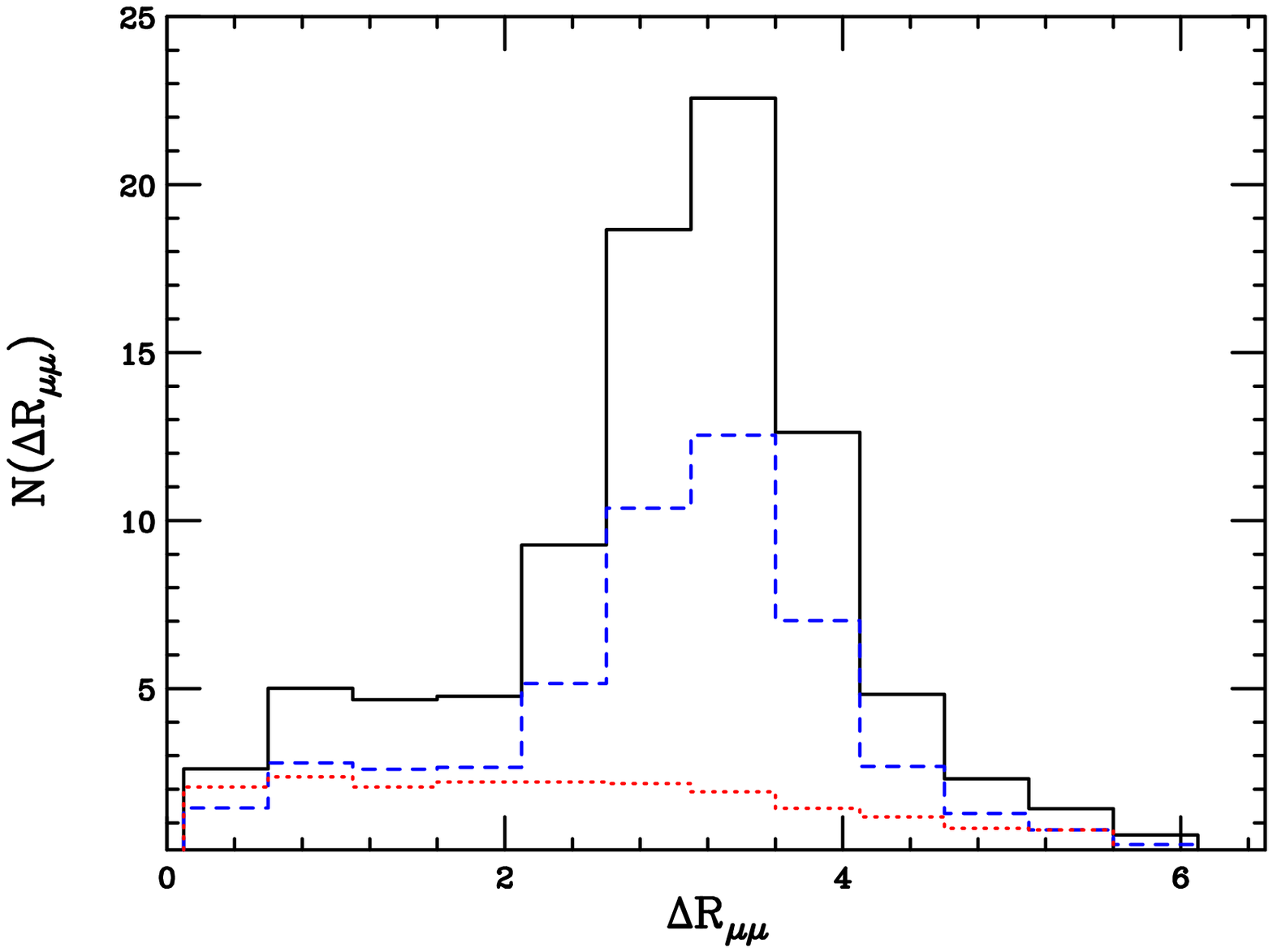}}}
\caption{Left: invariant mass of same-sign muons. Right: Invariant
  opening angle $\Delta R$ between the two hardest muons.
  The histograms refer to signal and backgrounds according to the convention in
Fig.~\ref{ptlep}}
\label{mlldr}
\end{figure}

\begin{figure}
\centerline{\resizebox{0.49\textwidth}{!}
{\includegraphics{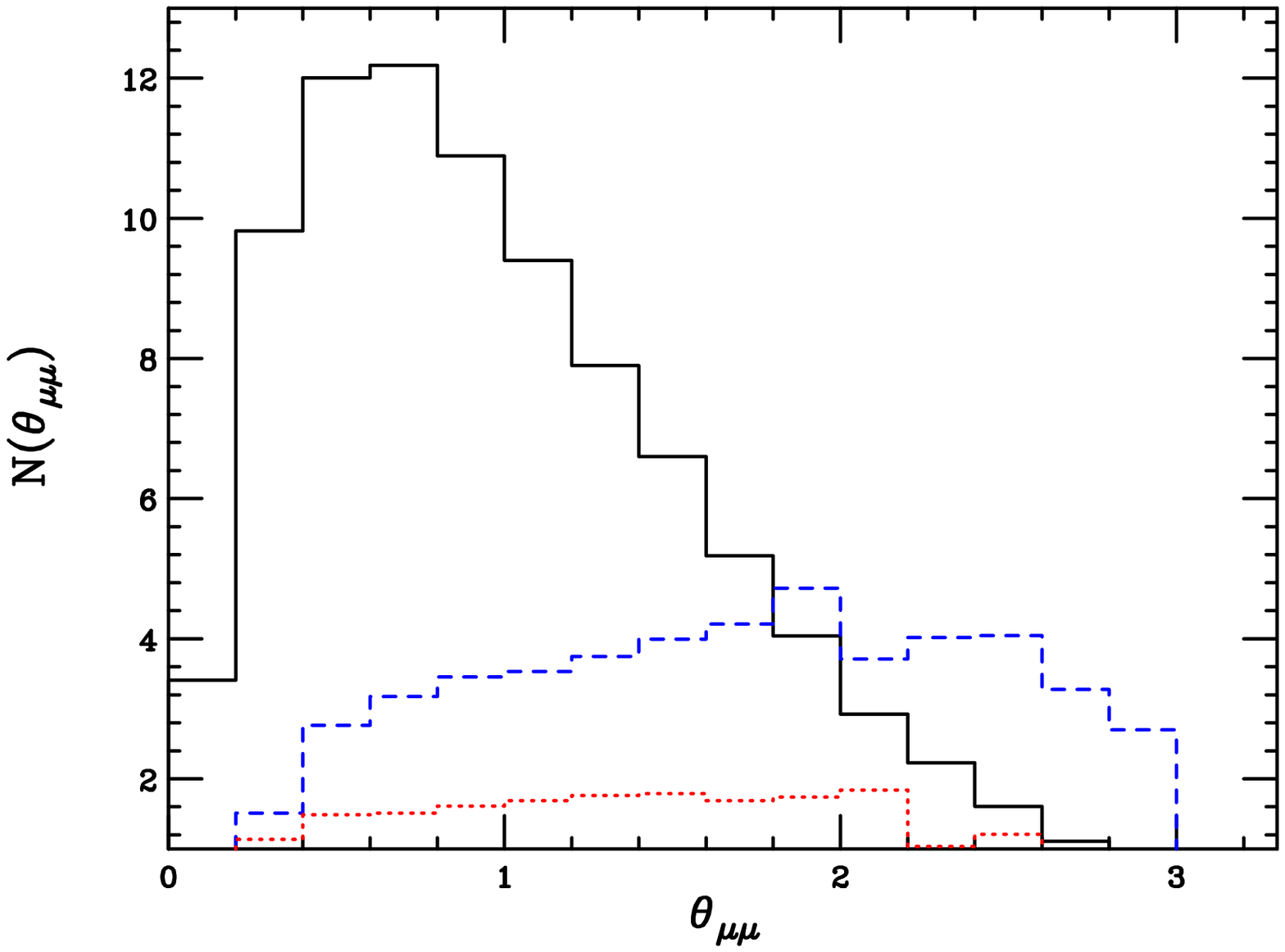}}%
\hfill%
\resizebox{0.49\textwidth}{!}{\includegraphics{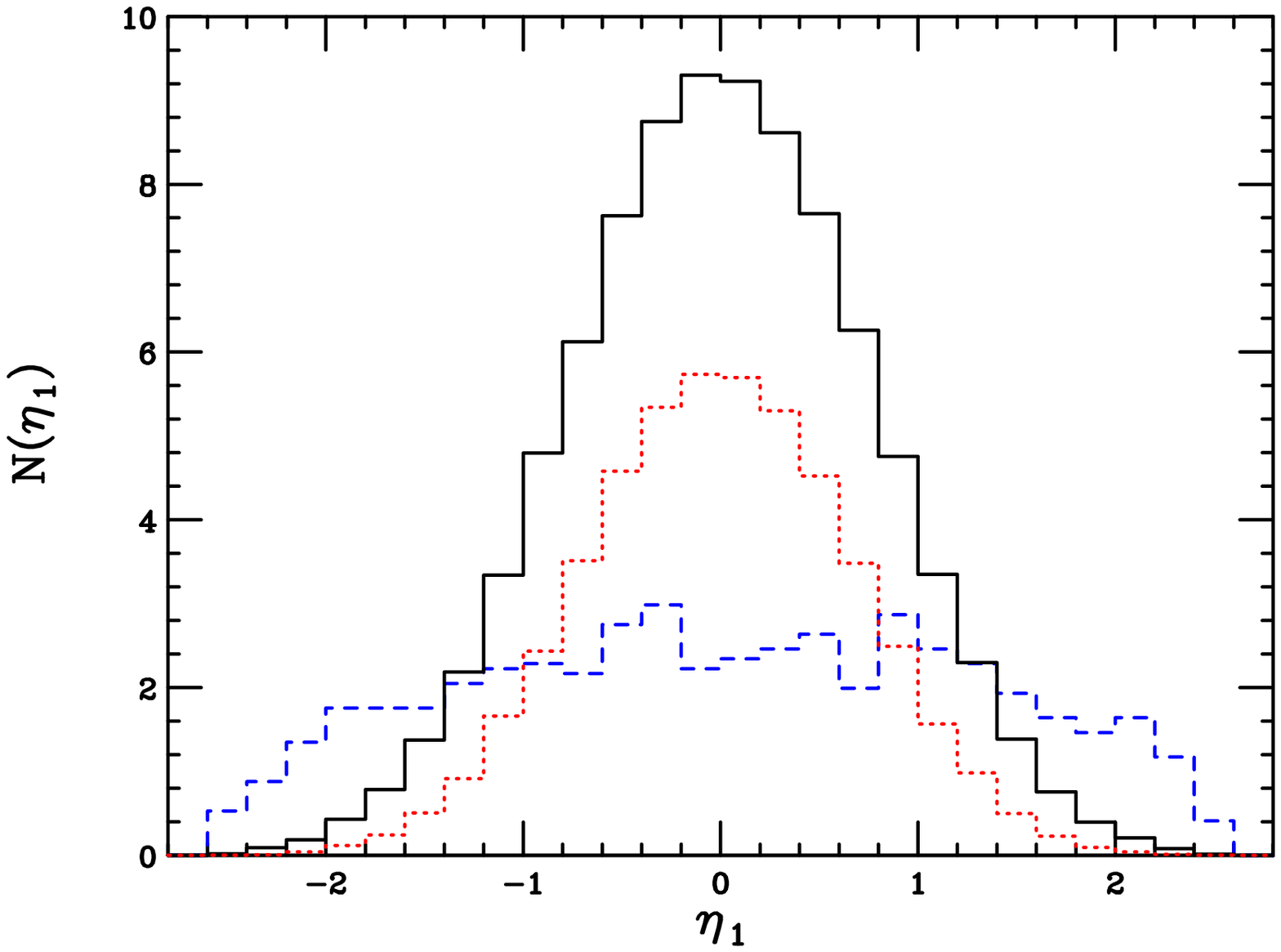}}}
\caption{Angle between the two hardest muons (left) and rapidity
  of the hardest muons. Histogram styles as in the previous figures.}
\label{thetaeta}
\end{figure}

\begin{figure}
\centerline{\resizebox{0.49\textwidth}{!}
{\includegraphics{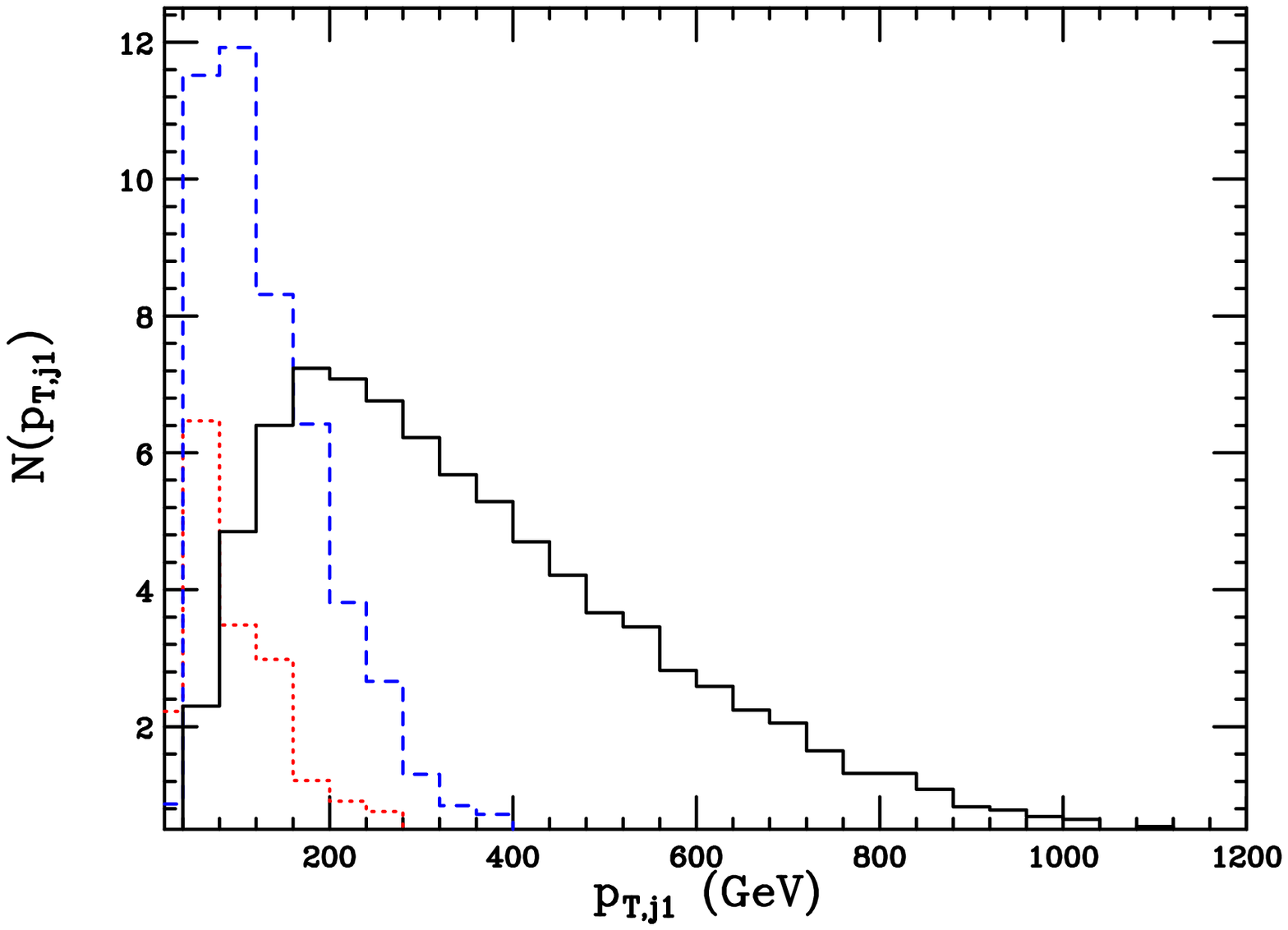}}%
\hfill%
\resizebox{0.49\textwidth}{!}{\includegraphics{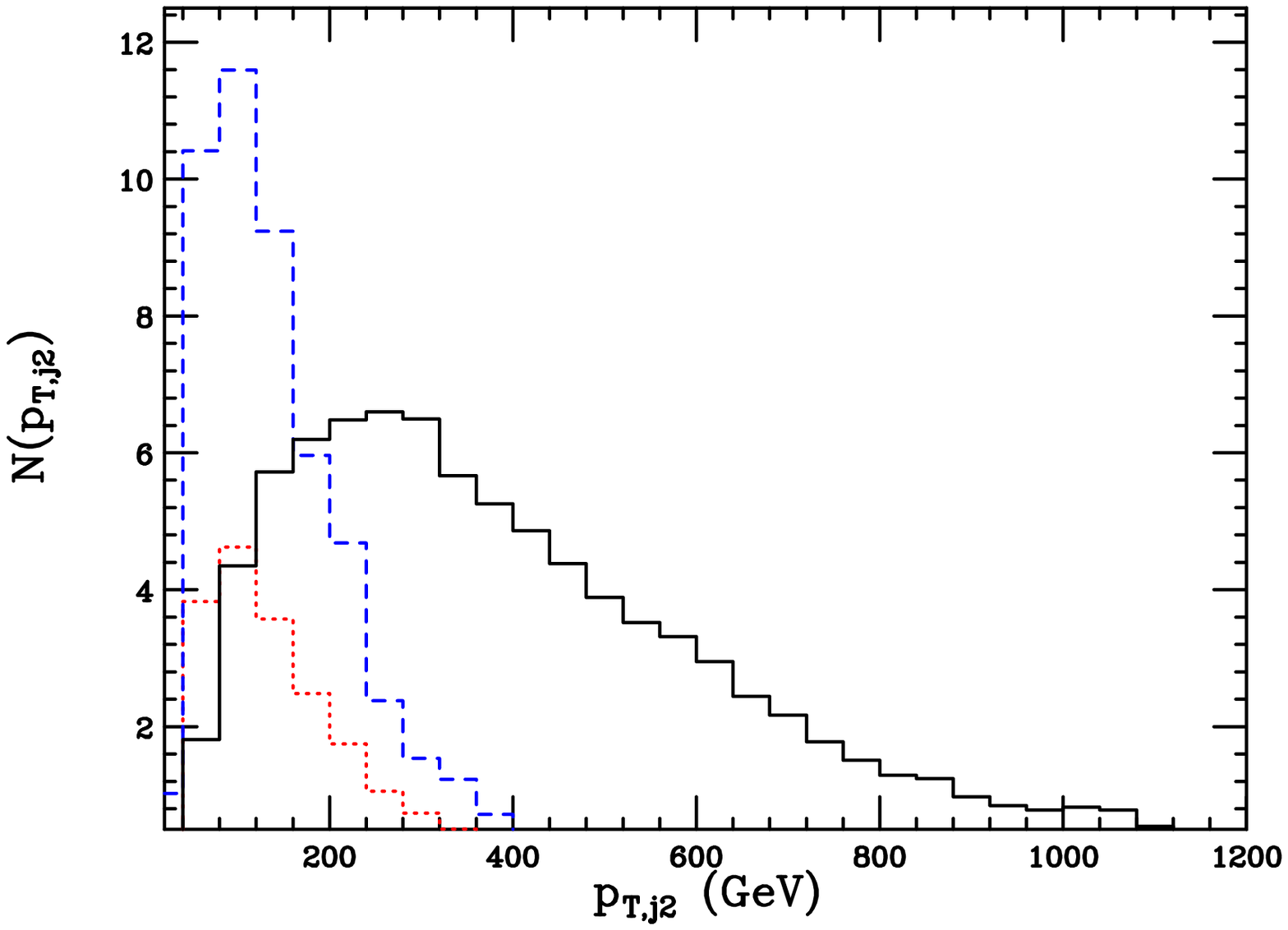}}}
\caption{First- (left) and second-jet (right) 
  transverse momentum, for the bilepton signal (solid) and Standard Model
  backgrounds (dots and dashes).}
\label{ptjet}
\end{figure}

\begin{figure}
\centerline{\resizebox{0.49\textwidth}{!}
{\includegraphics{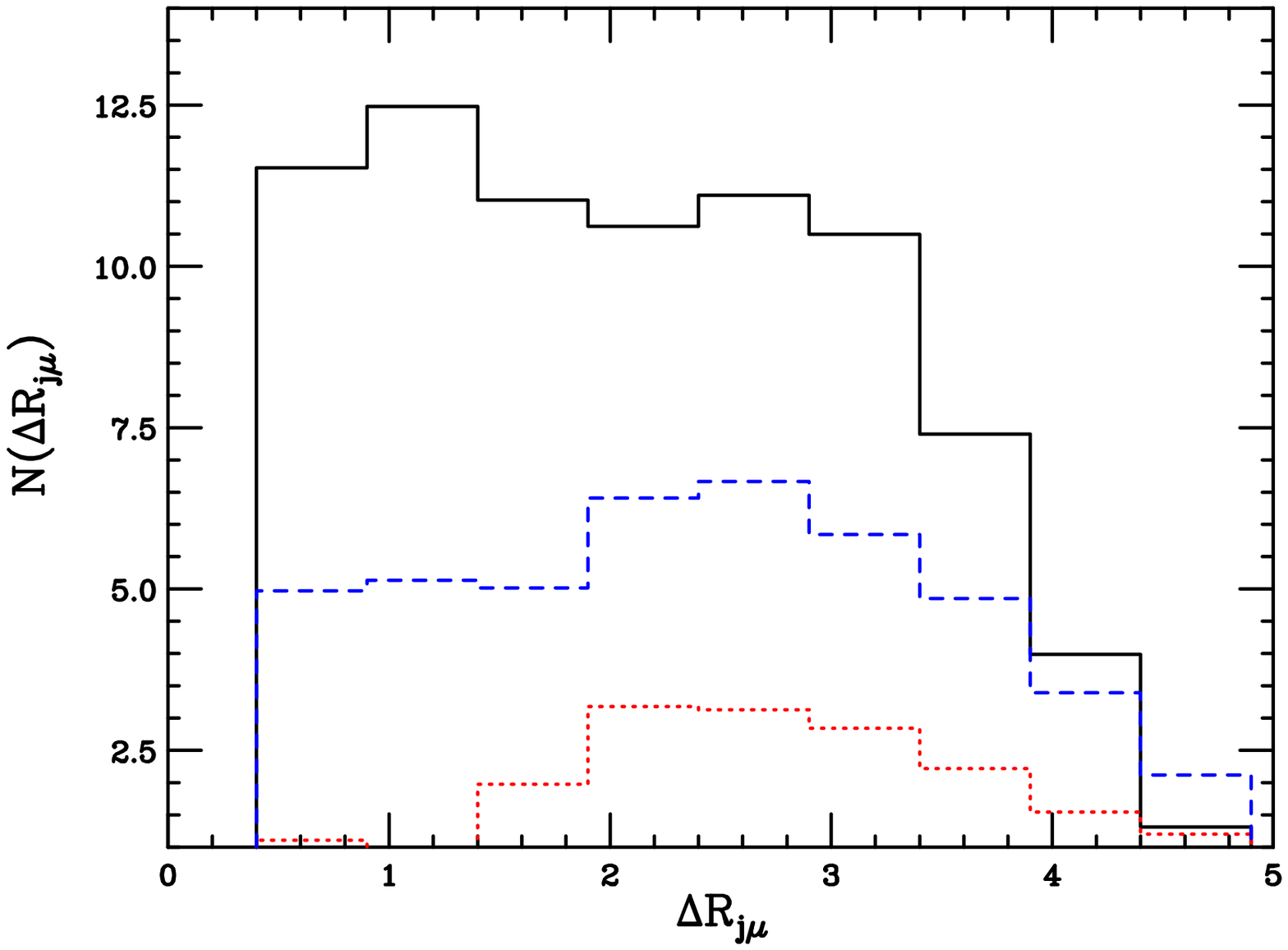}}%
\hfill%
\resizebox{0.49\textwidth}{!}{\includegraphics{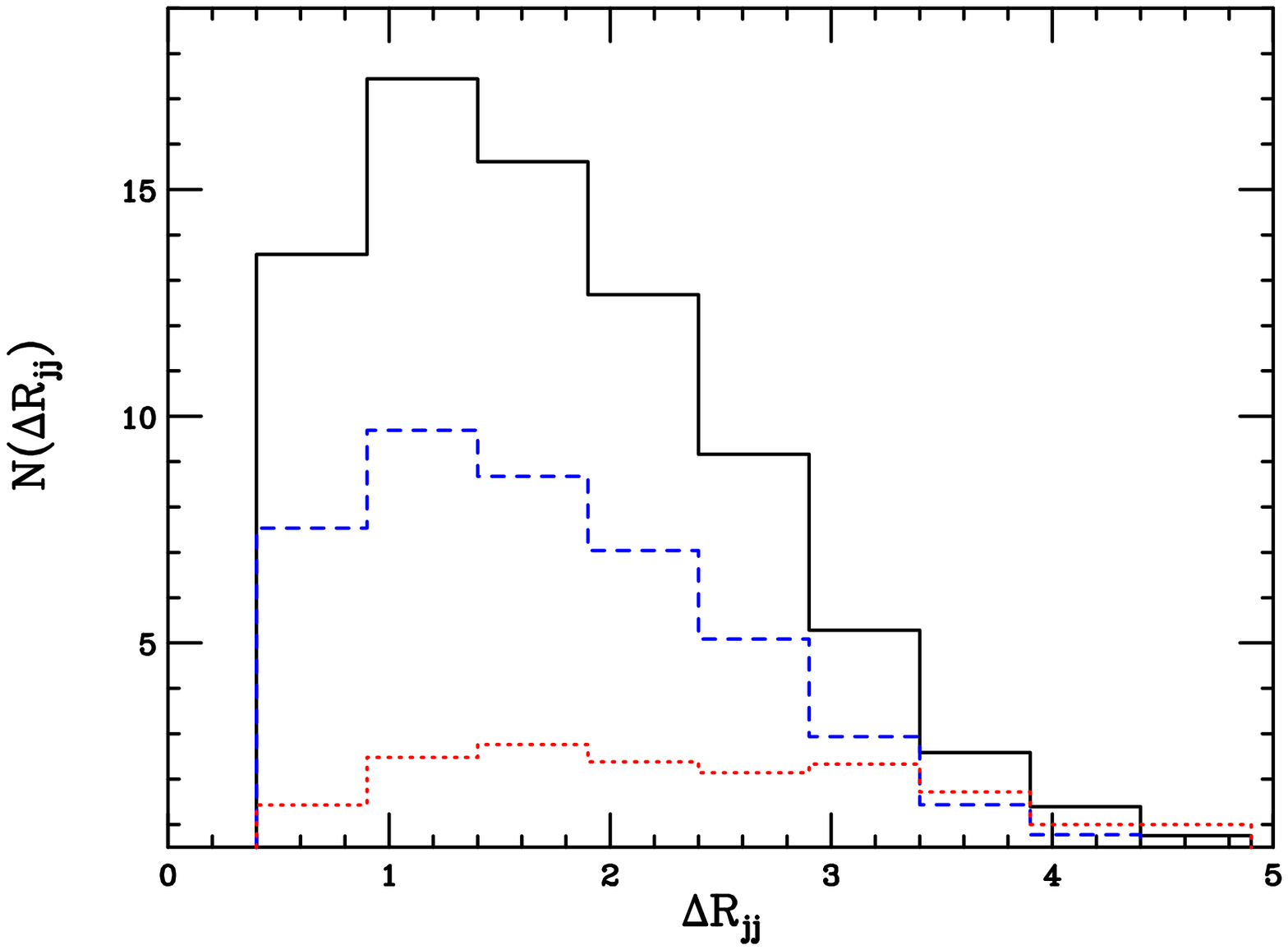}}}
\caption{As in Figs.~\ref{ptlep}--\ref{ptjet},
  but displaying the invariant opening angle between the hardest jet and the
  hardest muon (left) and the between first and second jet (right).}
\label{drjljj}
\end{figure}

In Figs.~\ref{ptlep}--\ref{drjljj} we present the distributions resulting
from our investigation: everywhere, our histograms are normalized in such
a way that the height of each bin, say $N(x)$, represents the foreseen
number of events at $x$ 
at FCC-$hh$.
Figure~\ref{ptlep} displays the spectra of the transverse momenta
of the hardest ($p_{T,1}$, left) and
next-to-hardest muon ($p_{T,2}$, right) according to the signal (\ref{dec1})
and backgrounds (\ref{back1}) and (\ref{back2}).
The muon $p_T$ spectra look quite similar:
the backgrounds are substantial only at low transverse momentum,
peak about $p_{T,1}\simeq 100$~GeV and $p_{T,2}\simeq 60$~GeV
and rapidly vanish at large $p_T$, in such a way that for $p_{T,1}>500$~GeV
and $p_{T,2}>300$~GeV only the signal survives. 
On the contrary, the
signal distributions are pretty broad and yield
some events up to $p_{T,1}\simeq 2$~TeV and $p_{T,2}\simeq 1.5$~TeV.
Requiring therefore muons with
transverse momenta above a few hundred GeV would therefore
help to discriminate the signal from the background:
this was  of course quite predictable, since, unlike the backgrounds,
our signal muons are 
indirectly related to the decay of a TeV-scale resonance, and therefore
a large $p_T$ is to be expected.

Regarding Fig.~\ref{mlldr}, where the invariant mass of same-sign muons
$M_{\mu\mu}$ (left) and
the invariant opening angle $\Delta R_{\mu\mu}$ (right)
between the two hardest muons
are presented, one can learn that, unlike the
backgrounds, whose $M_{\mu\mu}$ prediction peaks
at low values and becomes negligible above 500 GeV,
the signal yields a quite broad $M_{\mu\mu}$ spectrum, which is 
shifted towards large values and peaks at around 700 GeV.
As for $\Delta R_{\mu\mu}$, the signal is well above the backgrounds in the
whole range $0<\Delta R_{\mu\mu}<6$ and peaks at $\Delta R\simeq 3$.
The background $b_1$ exhibits a rather flat $\Delta R_{\mu\mu}$ distribution, 
while $b_2$ has a $\Delta R_{\mu\mu}$ spectrum with a shape similar to
the signal, though yielding a lower number of events in every bin.

Figure~\ref{thetaeta} displays the angle $\theta_{\mu\mu}$ (left)
between the two hardest
muons and the pseudorapidity $\eta_1$ (right) of the hardest one. 
For the purpose of 
$\theta_{\mu\mu}$, the signal spectrum is broad, peaks at 
$\theta_{\mu\mu}\simeq 0.8$ and is well above the backgrounds up to
$\theta_{\mu\mu}\simeq 1.8$. The backgrounds have a spectrum which
is rather flat, while for large angles,
say $\theta_{\mu\mu}>2$, the four-top background $b_2$ yields the highest rate
of events, above both signal and background $b_1$.
Concerning the pseudorapidity distribution, regardless of the
normalization, the shape of signal and background $b_1$ are
similar, with $b_1$ just predicting events with $|\eta|<1.8$.
The background $b_2$ exhibits instead a
quite flat $\eta$ distribution for $|\eta|<1.5$ and is even above the
signal for $1.5<|\eta|<2.5$.

In Figs.~\ref{ptjet} and \ref{drjljj} we instead explore the $b$-jet
properties, and in particular the transverse momenta $p_{T,j1}$ and
$p_{T,j2}$ of the first and second $b$-jet (Fig.~\ref{ptjet}) and the
invariant opening angles $\Delta R_{jj}$ and $\Delta R_{\mu j}$ between
the two hardest jets and between hardest jet and hardest muon
(Fig.~\ref{drjljj}). Overall, the comparison is similar to what
was observed for the leptonic observables.
While background jets typically peak at small transverse momenta,
about $p_T\simeq 60$~GeV $(b_1$) and  $p_T\simeq 1000$~GeV $(b_2$)
and vanish for $p_T>400$~GeV,
the spectra yielded by the $b$-jets originated from non-leptonic bilepton decays
are rather broad, dominate over the backgrounds above 300 GeV and give
meaningful event rates up to about 1.2 TeV.

Concerning the invariant opening angles plotted in Fig.~\ref{drjljj},
unlike the transverse momentum distributions, 
the shapes are more similar and the difference
among the spectra is mostly due to the overall normalization,
with the signal 
dominating over $b_1$ and $b_2$ 
for $\Delta R_{jl}<4.5$ and $\Delta R_{jj}<3.5$.
For larger opening angles the backgrounds tend to become competitive
with the signal and it is in fact $b_2$ which yields the highest rate for
$\Delta R_{jl}>4.5$ and $\Delta R_{jj}>3.5$.

\section{Conclusions}
We investigated the phenomenology of gauge bileptons $Y^{\pm\pm}$,
as predicted in the model \cite{PHF}, based on a 331 group
structure, paying special attention to their non-leptonic
decays in channels containing new TeV-scale quarks.
We chose two benchmarks points consistent with the current exclusion limits
on the searches for physics beyond the Standard Model and with the bilepton
mass estimate given in \cite{CF1}, and explored bilepton phenomenology at
LHC and at a future 100 TeV collider FCC-$hh$.

We found that
the cross sections are too low for any signal of non-leptonic decays 
of bileptons to be visible at LHC, at either 13 TeV and a luminosity of
300 fb$^{-1}$ or 14 TeV and ${\cal L}=3000~{\rm fb}^{-1}$.
Our first conclusion is therefore that, for the time being, the LHC can
detect bileptons only searching for same-sign lepton pairs,
along the lines of Refs.~\cite{CCCF1,CCCF2}, while the statistics are not 
sufficient to be sensitive to decays into new TeV-scale heavy quarks.
Nevertheless, non-leptonic bilepton decays could 
be observed at the FCC-$hh$ in a scenario wherein all
three new quarks are slightly lighter than the bilepton itself.
For this purpose, we explored decay chains
leading to four $b$-flavoured jets and
four muons through the primary production of bilepton pairs $Y^{++}Y^{--}$
decaying into quarks with charge $\pm 5/3$. 
The backgrounds due to 
two $t\bar t$ pairs, with all
top quarks decaying leptonically, or two  $b\bar b$ pairs and two
$Z$ bosons, eventually leading to two muon pairs were investigated as well.
We found that the signal can 
be discriminated at FCC-$hh$,
since it typically leads to events with muons and jets at large $p_T$
and same-sign muon pairs with high invariant mass
which dominate over the background in most phase space.

In summary, in the present study we have
demonstrated that it can be worth considering non-leptonic decays
of bileptons, but one
would need to wait for the very high-energy FCC-$hh$ for any realistic
search. Ideally, if the LHC were to have any bilepton signal in the
same-sign lepton-pair channel, then a 100 TeV collider could shed further
light on the  model, as it could reveal even the TeV-scale quarks
(see also Ref.~\cite{Frampton}),
which represent another striking feature of the bilepton scenario.

Of course, this paper and our previous work in \cite{CCCF1,CCCF2} can still be
extended according to several guidelines.
In fact, we plan to include in our investigation effects like detector
simulations, realistic tagging efficiencies for jets and leptons at LHC and
FCC-$hh$, as well as higher-order corrections to the partonic cross section
and parton distribution functions.
After accounting for such effects one could then give a realistic
estimate of the sensitivity of LHC and FCC-$hh$ to bileptons in the various
decay channels as a function of the most relevant model parameters.
Furthermore, as pointed out when discussing the backgrounds, the
four-top searches can be recast in terms of the bilepton model
from the LHC to the FCC-$hh$ energies.
Finally, it is obviously very interesting exploring
the primary production of TeV-scale quark pairs in the bilepton model and
their subsequent decays into bileptons, assuming that the heavy quarks 
are heavier than $Y^{\pm\pm}$.
This is in progress as well.

\section*{Acknowledgements}
We acknowledge Marco Zaro and Jack Araz for discussions on the \texttt{MadGraph}
simulation. We also thank Marianna Testa, Antonio Sidoti and Giacomo Polesello
for advices concerning the acceptance cuts on jets and leptons and 
the $b$-tagging efficiency.

\end{document}